\documentstyle[epsf,preprint,aps]{revtex}

\begin{document}


\title{Relations Among Correlation Functions in the High Temperature 
Phase of QCD with Broken SU(3)}

\author{Michael C. Birse$^{a,b}$, Thomas D. Cohen$^a$ and Judith A.
McGovern$^{a,b}$}

\address{$^a$ Department of Physics,\\
University of Maryland, College Park, MD, 20742-4111,USA\\ \ }

\address{$^b$ Theoretical Physics Group, Department of Physics and Astronomy,\\
University of Manchester, Manchester M13 9PL, U.K.}
\maketitle

\begin{abstract}
Group-theoretic arguments are used to determine the dependence of two-point
correlators of quark bilinears on the current quark masses. The leading
difference between $\pi$ and $\delta$ correlators is found to be of order
$m_s$ times a U(1)$_{\scriptscriptstyle A}$-violating correlator. These 
general arguments are consistent with Schaefer's observation that if
U(1)$_{\scriptscriptstyle A}$ violation persists to high enough temperatures
then the strange $\eta$ can be lighter than the non-strange one.
\vspace{20pt}
\end{abstract}

At high temperatures the SU($N_f$)$_{\scriptscriptstyle L}
\times$SU($N_f$)$_{\scriptscriptstyle R}$ chiral symmetry of QCD 
(with $N_f$ flavors of light quark) is believed to be broken only by the
nonzero current quark masses. The nature of this phase is of considerable
interest, in particular because of its relevance to ultrarelativistic heavy-ion
collisions and the early universe. One aspect of this phase that has received
recent interest\cite{shuryak,cohen,hatsuda,evans,schsh,bcm,schaefer} is the
role of the anomalously broken U(1)$_{\scriptscriptstyle A}$ symmetry at high
temperatures. Because of this anomaly, topological features of the QCD vacuum
split the $\eta'$ from the pion in the low-temperature phase\cite{thooft}.
This question, currently being studied using lattice-gauge simulations of
two-flavor QCD by a number of groups\cite{christ,boyd,bernard,kogut}, is
whether these topological features persist above the QCD phase transition.

The possible role of U(1)$_{\scriptscriptstyle A}$ symmetry breaking can be
studied by looking at correlation functions of composite operators constructed
from quark and gluon fields and comparing ones that are related by
U(1)$_{\scriptscriptstyle A}$ symmetry (and perhaps also
SU($N_f$)$_{\scriptscriptstyle L}\times$SU($N_f$)$_{\scriptscriptstyle R}$
symmetry). All observable manifestations of anomalous
U(1)$_{\scriptscriptstyle A}$ symmetry breaking should be reflected in the
behavior of these correlation functions. Shuryak has suggested that, as well
as the SU($N_f$)$_{\scriptscriptstyle L}\times$SU($N_f$)$_{\scriptscriptstyle
R}$ symmetry, U(1)$_{\scriptscriptstyle A}$ symmetry might also be restored in
the high temperature phase in the sense that no U(1)$_{\scriptscriptstyle
A}$-violating effects would be found in the correlation
functions\cite{shuryak}.

In a recent paper\cite{bcm}, we pointed out that in the limit of exact 
SU($N_f$)$_{\scriptscriptstyle L}\times$SU($N_f$)$_{\scriptscriptstyle R}$ 
chiral symmetry, this symmetry strongly constrains the ways in which
U(1)$_{\scriptscriptstyle A}$ violation can manifest itself in the
high-temperature phase. In particular it requires that all $n$-point
correlation functions of quark bilinears are invariant under
U(1)$_{\scriptscriptstyle A}$ transformations provided $n < N_f$. This means 
all two-point correlation functions are U(1)$_{\scriptscriptstyle A}$-invariant
for$ N_f>2$. This point had previously been noted within the context of
the instanton liquid model\cite{schsh} and under assumptions corresponding to
a dilute instanton gas\cite{hatsuda,evans}. The symmetry argument shows that
the result is general and does not depend on any particular picture for the
topological aspects of the QCD vacuum that generate the nonperturbative
U(1)$_{\scriptscriptstyle A}$ anomaly.

So far we have been discussing only the chiral limit of QCD, whereas in
reality the SU($N_f$)$_{\scriptscriptstyle
L}\times$SU($N_f$)$_{\scriptscriptstyle R}$ symmetry is explicitly broken by
the current quark masses. It is thus not immediately clear how relevant these
symmetry arguments are to a world with two light and one moderately light
quark flavors\cite{schaefer}. In the present note we extend our arguments of
Ref.\cite{bcm} to the case of broken chiral and SU(3) symmetries. The crucial
feature for this analysis is that sufficiently far above the phase transition,
or crossover region, it should be possible to make a perturbative
expansion of correlators in powers of the current quark masses. If the effects
of U(1)$_{\scriptscriptstyle A}$ violation persist up to such temperatures
then they can only be visible in correlators of quark bilinears in the ways
described here. These results should eventually be useful for interpreting
lattice simulations of three-flavor QCD with two light and one heavier
flavor of quark. They can also be used to understand, for example, the
results of instanton liquid calculations\cite{schaefer}.

In the restored phase of SU($N_f$)$_{\scriptscriptstyle
L}\times$SU($N_f$)$_{\scriptscriptstyle R}$ with zero quark masses, the only
non-vanishing vacuum correlators are SU($N_f$)$_{\scriptscriptstyle
L}\times$SU($N_f$)$_{\scriptscriptstyle R}$ singlets that are also even under
parity. Even when current quark masses are included, we can still use this
idea provided we restrict our consideration to temperatures that are
sufficiently far above the crossover region. In this situation we can make an
expansion of any correlator of quark bilinears in powers of the quark masses.
Each of the terms in such an expansion is itself a correlator of quark
bilinears. For example, at first order in $m_s$ the two-point correlator 
between quark bilinears with the quantum numbers of pions, 
$\langle\overline q(x)i\gamma_5\tau_1 q(x)\;\;\overline q(0)
i\gamma_5\tau_1 q(0)\rangle$,  contains
the three-point correlator
\begin{equation}
m_s\!\int\!d^4\!y\;\langle\overline q(x)i\gamma_5\tau_1 q(x)\;\;\overline q(0)
i\gamma_5\tau_1 q(0)\;\;\overline s(y)s(y)\rangle,
\end{equation}
where the expectation value is taken in an SU($N_f$)$_{\scriptscriptstyle
L}\times$SU($N_f$)$_{\scriptscriptstyle R}$-invariant vacuum. This correlator
can then be expressed in terms of SU($N_f$)$_{\scriptscriptstyle
L}\times$SU($N_f$)$_{\scriptscriptstyle R}$ singlets as described in
Ref.\cite{bcm}. Similarly for a correlator of $n$ bilinears, a term of 
$m$-th order in the quark masses can be related to singlets constructed out
of $n+m$ bilinears.
In the case of three flavors, this analysis leads to some nontrivial
conclusions which are relevant to the study of U(1)$_{\scriptscriptstyle A}$
violation in high temperature QCD.

To derive such relations, it is convenient to consider the partition function
for QCD in the presence of scalar and pseudoscalar sources $S_a(x)$ and 
$P_a(x)$, where $a$ runs from 0 to $N_f^2-1$:
\begin{equation}
Z=\int\!{\cal D}q\,{\cal D}\overline q\,{\cal D}A\,\exp\left[-\int\!d^4\!x\,
\Bigl({\cal L}_{\rm QCD}(x)-S_a(x)\overline q(x)\lambda_a q(x)
-P_a(x)\overline q(x)i\gamma_5\lambda_a q(x)\Bigr)\right]
\end{equation}
where the path integral is over fields that are (anti-)periodic in Euclidean
time. Connected correlation functions of the quark bilinears can be expressed
as derivatives of $\ln Z[S_a,P_a]$ with respect to the sources. For example, 
the correlator in the pion channel is
\begin{equation}
\langle\overline q(x)i\gamma_5\tau_1 q(x)\;\;\overline q(0)
i\gamma_5\tau_1 q(0)\rangle=\left.{\delta^2\ln Z[S_a,P_a]\over\delta P_1(x)
\delta P_1(0)}\right|_{\rm phys}.
\label{mat4}
\end{equation}
The subscript indicates that the sources are set to their physical values.
If we omit the quark mass terms from ${\cal L}_{\rm QCD}$ then the derivatives
of $\ln Z[S_a,P_a]$ at $S_a=P_a=0$ give the correlators in the chiral limit.
Correlators in the case of explicit symmetry breaking can be obtained by
setting the appropriate combinations of scalar sources to equal the current
quark masses.

For vanishing current quark masses chiral symmetry requires that $\ln
Z[S_a,P_a]$ be invariant under SU($N_f$)$_{\scriptscriptstyle
L}\times$SU($N_f$)$_{\scriptscriptstyle R}$ transformations. 
It can thus be expressed in terms of the group invariants.  These are 
conveniently constructed from the following structures:
\begin{equation}
{\bf \Sigma}=\sum_{a=0}^{N_f^2-1} (S_a+i P_a)\lambda_a\qquad\hbox{and}\qquad
{\bf \Sigma}^\dagger=\sum_{a=0}^{N_f^2-1} (S_a-i P_a)\lambda_a
\label{mat1}
\end{equation}
which transform respectively under the representations $(N_f,\overline N_f)$
and $(\overline N_f,N_f)$ of SU($N_f$)$_{\scriptscriptstyle
L}\times$SU($N_f$)$_{\scriptscriptstyle R}$: ${\bf \Sigma\to 
U}_{\scriptscriptstyle L}{\bf \Sigma U}_{\scriptscriptstyle R}^\dagger$ and 
${\bf \Sigma^\dagger\to U}_{\scriptscriptstyle R}{\bf \Sigma^\dagger 
U}_{\scriptscriptstyle L}^\dagger$. Under a U(1)$_{\scriptscriptstyle A}$
transformation $\bf \Sigma$ transforms as ${\bf \Sigma}\to e^{i\theta}
{\bf \Sigma }$, and under a parity operation, 
as ${\bf \Sigma}\to {\bf \Sigma}^\dagger$.

As discussed in Ref.\cite{bcm}, these singlets can be constructed in two
distinct ways. One is to take equal numbers of $\bf \Sigma$'s and $\bf
\Sigma^\dagger$'s, coupled up to a singlet.  Examples are ${\rm Tr}\,
({\bf \Sigma}_1^\dagger {\bf \Sigma}_2)$, ${\rm Tr}\,({\bf \Sigma}_1^\dagger 
{\bf \Sigma}_2 {\bf \Sigma}_3^\dagger {\bf \Sigma}_4)$, ${\rm Tr}\,
({\bf \Sigma}_1^\dagger {\bf \Sigma}_2) {\rm Tr}\,({\bf \Sigma}_3^\dagger 
{\bf \Sigma}_4)$, etc.\ (where ${\bf \Sigma}_i \equiv {\bf
\Sigma}(x_i)$). All of these are not only chiral invariants; they are obviously
U(1)$_{\scriptscriptstyle A}$-invariant as well.

The only other way of obtaining a singlet is to couple $N_f$ $\bf \Sigma$'s or
$N_f$ $\bf \Sigma^\dagger$'s together in an antisymmetric way. This produces two
singlets:
\begin{equation}
{1\over (N_f) !}\epsilon_{ijk\ldots p} \epsilon_{i'j'k'\ldots p'} 
({\bf \Sigma}_1)_{ii'}({\bf \Sigma}_2)_{jj'}({\bf \Sigma}_3)_{kk'}\ldots
({\bf \Sigma}_{\scriptscriptstyle N_f})_{pp'}
\label{det}
\end{equation}
and the analogous expression with $\bf \Sigma\to \Sigma^\dagger$. (All indices
run from 1 to $N_f$.)  (For identical $\bf \Sigma$'s, these terms are just
$\det {\bf \Sigma}$ and $\det {\bf \Sigma^\dagger}$.)  By parity, only the sum
has a nonvanishing vacuum expectation value. Unlike the invariants
constructed out of equal numbers of $\bf \Sigma$'s and $\bf \Sigma^\dagger$'s,
this is not U(1)$_{\scriptscriptstyle A}$-invariant.  Further chiral singlet,
U(1)$_{\scriptscriptstyle A}$-violating terms may be obtained by coupling, for
instance, $N_f+1$ $\bf \Sigma$'s and an $\bf \Sigma^\dagger$, etc.

Provided that the temperature considered is above the
phase transition and so $\ln Z[S_a,P_a]$ has an analytic dependence on the
sources, we can express $\ln Z[S_a,P_a]$ as a linear combination of 
SU($N_f$)$_{\scriptscriptstyle L}\times$SU($N_f$)$_{\scriptscriptstyle R}$ 
invariants. Each term must have the form
\begin{equation}
\int d^4\!x_1\ldots d^4\!x_m f(x_1\ldots x_m) I_m(x_1\ldots x_m)
\end{equation}
where $I_m$ is one of the invariants discussed above.
The weighting function must be translationally invariant and 
symmetric under interchange of the arguments $x_i$.  Such a term then gives a
contribution to the correlator (\ref{mat4}) 
of the form $g(x)(d^2J_m/dP_1^2)_{\rm phys}$ where
$g(x)$ is an invariant function and $J_m$ is the space-time-independent
invariant obtained by setting all the of the sources in $I_m$ to constants.
Of the $J_m$, $N_f+1$ are the
independent invariants of SU($N_f$)$_{\scriptscriptstyle
L}\times$SU($N_f$)$_{\scriptscriptstyle R}$, conveniently taken as  
${\rm Tr}[({\bf \Sigma}^\dagger{\bf \Sigma})^n]$, $n\le N_f$, and 
$(\det{\bf \Sigma}+\det{\bf \Sigma^\dagger})$, and all the others are powers
and products of these.

In the chiral limit, of course, all the sources $S_a$ and $P_a$ vanish.  Thus 
only the $m$th order invariants contribute to the $m$-point correlators.
Hence, as shown in Ref.~\cite{bcm}, all $m$-point correlators for $m<N_f$ are
U(1)$_{\scriptscriptstyle A}$-invariant since the lowest 
U(1)$_{\scriptscriptstyle A}$-violating invariant is the determinant term.
However, by studying the actual forms of the invariants for a given $N_f$, one
can make predictions about the quark-mass-dependence of correlators.  Here
we will be concerned with two-point correlators in the physically interesting
case of three flavors, with $m_u$ and $m_d$ light (and taken to be equal),
and $m_s$ not so light. 
Using the same notation as in Ref.\cite{bcm} we can denote the scalar quark
bilinears $\overline q(x)\lambda_a q(x)$ by $\xi_a(x)$, where $a=0$ refers
to the flavor singlet and $a=1,\ldots N_f^2-1$ to the rest. Similarly we
use $\phi_a(x)$ to denote the corresponding pseudoscalars.
As a shorthand, we will use meson names for these quark bilinears:
the pions and kaons,
the scalar pions $\delta$, ($\xi_1\ldots\xi_3$), and scalar kaons $\kappa$, 
($\xi_4\ldots\xi_7$), the non-strange and strange scalar singlets, 
$\sigma=(\sqrt 2 \xi_0+\xi_8)/\sqrt 3$ and 
$\zeta=(\xi_0-\sqrt 2 \xi_8)/\sqrt 3$, and the pseudoscalars, 
$\eta_n$ and $\eta_s$, which are the
corresponding combinations of $\phi_0$ and $\phi_8$.  Of course the last four
need not be mass eigenstates and there need not be bound states in any channel.
There are nonvanishing vacuum expectation values for $\sigma$ and $\zeta$ for
which the sources are just $m_u$ and $m_s$.   We can then make the 
following observations.

First, as SU(2)$_{\scriptscriptstyle L}\times$SU(2)$_{\scriptscriptstyle R}$ is
a subgroup of 
SU(3)$_{\scriptscriptstyle L}\times$SU(3)$_{\scriptscriptstyle R}$, if we
restrict ourselves to the sources for the non-strange bilinears and $\zeta$
(which is
U(2)$_{\scriptscriptstyle L}\times$U(2)$_{\scriptscriptstyle R}$-invariant),
the three-flavor invariants can be written in terms of the two-flavor
invariants and $\zeta$.  Thus if two correlators are equal in the
$N_f=2$ chiral limit, their difference is only of order $m_u^2$ with realistic
masses.

Second, for any source except for those for $\sigma$ and $\zeta$,  the first
derivative of an irreducible invariant with respect to that source vanishes
by conservation of flavor quantum numbers and parity when the physical values
of the sources are substituted. Thus the second derivative of a product of
irreducible invariants vanishes  unless both derivatives act on the same
factor.  The only group structures which arise in the differences between
two-point correlators of these bilinears are those  which come from the four
irreducible invariants, and all higher terms simply alter the invariant
coefficients.  By explicit calculation, differences between correlators can be
found which are of order $m_u^2$ and therefore small even if $m_s$ is large. 
Others can be found which will be of order $m_s$ if the anomaly persists, but
$m_u^2$ otherwise. These are therefore sensitive tests of
U(1)$_{\scriptscriptstyle A}$ restoration.  Denoting the
correlator in (\ref{mat4}) as $\Pi_\pi$ and others accordingly, and with
$\gamma$ representing the strength of the anomaly, the relations are
\begin{eqnarray}
\Pi_\pi\! &=&\! \Pi_\sigma+O(m_u^2)\nonumber \\
\Pi_\delta\! &=&\! \Pi_{\eta_n}+O(m_u^2)\nonumber\\
{\textstyle \frac 1 2 }(\Pi_{\scriptscriptstyle K}+\Pi_\kappa)\! &=&\! 
\Pi_{\eta_s}+O(m_u^2)\nonumber\\
\Pi_\pi\! &=&\! \Pi_\delta+O(\gamma m_s)+O(m_u^2)\nonumber\\
\Pi_{\eta_n}\! &=&\! \Pi_{\eta_s}+O(\gamma m_s)+O(m_s^2)
\end{eqnarray}
In addition the mixed $\eta_n\eta_s$ correlator is of order $\gamma m_u$, so 
even if the anomaly persists the mixing is small.  (This contrasts with the
low-temperature situation where the mass eigenstates are close to $\eta_0$
and $\eta_8$.)

The difference between the correlators in the pion
and $\delta$ (scalar pion) channels is widely used to study
U(1)$_{\scriptscriptstyle A}$ violation in two-flavor
QCD\cite{christ,boyd,bernard,kogut} since both correlators can be determined
from lattice calculations without need for quark-line disconnected pieces.
In the three-flavor case we find that the leading term in the difference
between these correlators is of order $m_s$ times a three-point 
U(1)$_{\scriptscriptstyle A}$-violating correlator. Corrections
to this are of order $m_u^2$. Hence due to SU(2)$\times$SU(2) symmetry the
$\pi-\delta$ difference remains a good test of whether
U(1)$_{\scriptscriptstyle A}$ violation persists at high temperatures.

There has also been recent interest in the behavior of the SU(2) singlet 
pseudoscalars ($\eta$ and $\eta'$) at high 
temperatures\cite{kapusta,huang,schaefer}. However the correlators in both of
these channels require quark-line disconnected pieces. We find that the 
non-strange and strange $\eta$'s decouple, 
as noted by Schaefer in the instanton-liquid model\cite{schaefer}. Moreover
the correlator for the non-strange $\eta$ is the same as that for the $\delta$
while that for the strange $\eta$ is equal to the average of the kaon and
$\kappa$ (scalar kaon) correlators. Corrections to these relations are also of
order $m_u^2$. They provide a way to determine the correlators in
the $\eta$ channels without the need to evaluate disconnected pieces.

Finally, the leading term in the difference between the strange and non-strange
$\eta$'s is of order $m_s$ times an anomalous three-point correlator, with 
corrections of order $m_s^2$.  If the effect of the anomaly has the same sign
above the transition temperature as below, the first term
has a sign corresponding to a larger screening mass in the non-strange channel.
This rather surprising result was first noted by Schaefer\cite{schaefer}.
However the size of the corrections means that for
this effect to be visible the nonperturbative effects of the anomaly are
likely to have to survive well above the crossover region.

This work was  supported in part by the US Department of Energy
under grant no. DE-FG02-93ER-40762, and in part by the UK EPSRC. M.C.B.\ and
J.McG.\ would like to thank the TQHN group at the University of Maryland for
their generous hospitality.


\begin{thebibliography}{99}

\bibitem{shuryak} E. V. Shuryak, Comments Nucl.\ Part.\ Phys.\ {\bf 21}, 
235 (1994); report hep-ph/9503427 (1995).
\bibitem{cohen} T. D. Cohen, Phys.\ Rev.\ {\bf D54}, 1867 (1996).
\bibitem{hatsuda} S. H. Lee and T. Hatsuda, Phys.\ Rev.\ {\bf D54},   1871
(1996).
\bibitem{evans} N. Evans, S. D. H. Hsu and M. Schwetz, Phys.\ Lett.\ {\bf B375} 
262, (1996).
\bibitem{schsh} T. Schaefer and E. V. Shuryak, Phys.\ Rev.\ {\bf D54}, 1099
(1996).
\bibitem{bcm} M. C. Birse, T. D. Cohen and J. A. McGovern, University of
Maryland report 97-014 (1996), hep-ph/9608255, Phys.\ Lett.\ {\bf B} (to be
published).
\bibitem{schaefer} T. Schaefer, University of Washington report
DOE/ER/40561-280, INT96-00-140 (1996), hep-ph/9608373.
\bibitem{thooft} G. 't Hooft, Phys.\ Rev.\ Lett.\ {\bf 37}, 8 (1976); 
Phys.\ Rev.\ {\bf D14}, 3432 (1976).
\bibitem{christ} S. Chandrasekharan and N. Christ, Nucl.\ Phys.\ B 
(Proc.\ Suppl.) {\bf 47}, 527 (1996).
\bibitem{boyd} G. Boyd,  F. Karsch, E. Laermann and M. Oevers, University of
Bielefeld report BI-TP 96/27 (1996), hep-lat/9607046.
\bibitem{bernard} C. Bernard {\it et al.}, Utah University report UU-HEP 96/5
(1996), hep-lat/9608026; Utah University report UU-HEP 96/8 (1996), 
hep-lat/9611031.
\bibitem{kogut} J. B. Kogut, J.-F. Lagae and D. K. Sinclair, Argonne report
ANL-HEP-CP-96-62 (1996), hep-lat/9608128.
\bibitem{kapusta} J. Kapusta, D. Kharzeev and L. McLerran, Phys.\ Rev.\ {\bf 
D53}, 5028 (1996).
\bibitem{huang} Z. Huang and X.-N. Wang, Phys.\ Rev.\ {\bf D53}, 5034 (1996).

\end{thebibliography}
\end{document}